\newcommand{\aap}{{\em A.\&A.{\rm}}}

\newcommand{\aj}{{\em A.~J.{\rm}}}
\newcommand{\apj}{{\em Ap.~J.{\rm}}}
\newcommand{\apjl}{{\em Ap.~J. Letters{\rm}}}
\newcommand{\apjs}{{\em Ap.~J. Suppl.{\rm}}}

\newcommand{\mnras}{{\em M.N.R.A.S.{\rm}}}
\newcommand{\pasp}{{\em P.A.S.P.{\rm}}}

\documentclass{iaus}
\usepackage{graphicx}

\title[Galaxy Groups and $H_0$ from Lenses]{Galaxy Groups Associated with
Gravitational Lenses and $H_0$ from B1608+656}

\author[Fassnacht et al.]%
{Chris Fassnacht$^1$,
Lori Lubin$^1$, John McKean$^1$, Roy Gal$^1$,\break
Gordon Squires$^2$, Leon Koopmans$^3$, Tommaso Treu$^4$, Roger Blandford$^5$,
\and David Rusin$^6$}

\affiliation{$^1$Department of Physics, UC Davis, 1 Shields Ave.,
Davis, CA 95616, USA email: fassnacht, lubin, mckean, gal@physics.ucdavis.edu\\
[\affilskip]
$^2$Spitzer Science Center, California Institute of Technology,
Mail Code 220-6,
1200 East California Boulevard,
Pasadena, CA 91125, USA email: squires@ipac.caltech.edu\\[\affilskip]
$^3$Kapteyn Institute, PO Box 800, 9700AV Groningen, The Netherlands
email: koopmans@astro.rug.nl\\[\affilskip]
$^4$Department of Physics and Astronomy, UCLA, Box 951547, Knudsen Hall, 
Los Angeles, CA 90095, USA email: ttreu@astro.ucla.edu\\[\affilskip]
$^5$KIPAC, Stanford University, 2575 Sand Hill Road, Menlo Park, CA 94025, USA
email: rdb3@stanford.edu\\[\affilskip]
$6$Department of Physics and Astronomy, University of Pennsylvania, 
Philadelphia, Pennsylvania 19104, USA email: drusin@astro.upenn.edu
}

\pubyear{2004}
\volume{225}
\pagerange{1--8}
\date{?? and in revised form ??}
\jname{Impact of Gravitational Lensing on Cosmology}
\editors{Mellier, Y. \& Meylan,G. eds.}
\begin{document}

\maketitle

\begin{abstract}
Compact groups of galaxies recently have been discovered in
association with several strong gravitational lens systems.  These
groups provide additional convergence to the lensing potential and
thus affect the value of $H_0$ derived from the systems.  Lens
system time delays are now being measured with uncertainties of only a
few percent or better.  Additionally, vast improvements are being made
in incorporating observational constraints such as Einstein ring
structures and stellar velocity dispersions into the lens models.  These
advances are reducing the uncertainties on $H_0$ to levels at which
the the effects of associated galaxy groups may contribute
significantly to the overall error budget.  We describe a dedicated
multiwavelength program, using Keck, HST, and Chandra, to find such
groups and measure their properties.  We present, as a case study,
results obtained from observations of the CLASS lens system B1608+656
and discuss the implications for the value of $H_0$ derived from this
system.
\end{abstract}

\firstsection 
\section{Introduction}

The determination of the Hubble Constant from gravitational lenses,
first described by Refsdal (1964), is an elegant method that is
completely independent from traditional distance-ladder techniques.
Recently the uncertainties on lens-based determinations of $H_0$ have
been dropping.  Intensive monitoring campaigns have led to precise
time-delay measurements (e.g., Kundi\'c et al.\ 1997c; Biggs et al.\
1999; Fassnacht et al.\ 2002; Burud et al. 2002; Hjorth et al. 2002).
Similarly, high angular resolution imaging from the {\em Hubble Space
Telescope} (HST) or Very Long Baseline Interferometry (VLBI) observations,
measurements of the stellar velocity dispersions of lensing
galaxies, and improvements in modeling codes have reduced the often
large uncertainties due to the lens model (e.g., Treu \& Koopmans 2002;
Rusin et al.\ 2002; Cohn et al.\ 2001).  With uncertainties
approaching 10\% or less for individual lens systems, it is now
necessary to consider other small sources of error in the
determination of $H_0$.  In this paper we consider the effect
of the environment of the lensing galaxy, namely the contribution of
galaxy groups to the lensing potential.

The presence of groups associated with gravitational lenses is not
unexpected.  In the local Universe, small groups provide the most
common galaxy environment (e.g., Turner \& Gott 1976; Geller \& Huchra
1983; Tully 1987; Ramella et al.\ 1989).  Additionally, local
elliptical galaxies are preferentially found in dense environments
(e.g., Dressler 1980; Zabludoff \& Mulchaey 1989).  Since most lenses
are early-type galaxies, we might expect to find them in groups if the
morphology-density relationship continues to moderate redshifts.
Theoretical studies have predicted that 25\% (Keeton et al.\ 2000) or
more (Blandford et al.\ 2001) of lenses should reside in compact
groups.  The presence of groups associated with lenses has important
implications for $H_0$ determinations.  The additional convergence
($\kappa$) provided by the group mass will lead to a value of $H_0$
that is too high if the group contribution is not properly taken into
account, i.e.,
$$
H_{0,true} = H_{0,meas} (1 - \kappa_{group}).
$$
The problem is that the standard lens observables (e.g., image
positions and flux ratios) do not provide any indication of the
presence of a group; this is the famous ``mass-sheet degeneracy''
(e.g., Falco et al.\ 1985).  

Keeton and Zabludoff (2004) have performed simulations that examine
the effect of groups on the measurement of lens properties.  Their
results on $H_0$ confirm the effect of the mass-sheet degeneracy.
That is, the value of $H_0$ derived from a lens system {\em without
taking into account the presence of the associated group} is
systematically high.  Although the large range of values that their
simulations produce may be due to the details of their simulations
(see comments by Chris Kochanek at this meeting), the fact remains
that if the group is not included in the lens model the resulting
value of $H_0$ will be biased high.  Also, the effect of the group on
the final value of $H_0$ depends on the individual characteristics of
the group.  Thus, it appears that it will not be possible to apply a
statistical correction for the possible presence of groups associated
with lenses.  The simulations also show that the input values are
recovered more accurately when the group is modeled as individual
halos rather than an overall smooth mass distribution.  Therefore,
it is important to search for groups that may be associated with
lens systems and, if such groups are found, measure their properties.

\section{A dedicated survey for lens-associated groups}

The mass-sheet degeneracy has long been acknowledged as a problem with
using lenses to determine $H_0$ (e.g., Falco et al.\ 1985).
Therefore, we have been conducting a systematic search for galaxy
groups associated with gravitational lenses.  The goal of the program
is to determine the group properties and correct for the group
contribution to the overall lens potential.  For a given lens system,
the initial observations consist of multi-color ground-based images of
the field containing the lens system.  These images are used to select
targets for spectroscopy.  High-priority targets are those which (1)
have colors similar to those expected for early-type galaxies at the
lens redshift, and (2) have small projected offsets from the lens
system (regardless of color).  These targets are then
spectroscopically observed with LRIS (Oke et al.\ 1995) and ESI
(Sheinis et al.\ 2002) at the Keck Telescopes.  The main instrument
used is LRIS because of its multislit capability.  In order to
maximize the number of slits on any given mask, some low-priority
targets are also included.  The spectroscopic observations are used to
produce a redshift distribution for the field which is then searched
for spikes that may correspond to groups or clusters.  Group
candidates are followed up with multi-wavelength observations such as
{\em Chandra} imaging to search for hot intragroup gas and HST imaging
to assess the properties of confirmed group galaxies.

\section{CLASS B1608+656}

The B1608+656 lens system remains the only four-image system for which
all three independent time delays have been measured to high precision
(Fassnacht et al.\ 2002).  New modeling incorporates the Einstein ring
seen in HST/WFPC2 images as well as the stellar velocity dispersion of
the lensing galaxy.  The new model leads to a value of $H_0 =
75^{+7}_{-6}$~km/s/Mpc (Koopmans et al.\ 2003).  Furthermore, a
20-orbit multi-band observation of the system with the Advanced Camera
for Surveys on HST has been acquired.  These deep imaging data will be
combined with new software to incorporate fully the Einstein ring data
into the lens model.  With this wealth of observational information,
the B1608+656 system is a natural target for the group-search program.

In order to choose targets for spectroscopic observations, we imaged
the B1608+656 field with the Palomar 60-Inch telescope.  Images
were obtained in three Gunn bands: $g$, $r$, and $i$.  Catalogs were
created from the images using the SExtractor package (Bertin \&
Arnouts 1996) and the prioritized target lists were created from these
catalogs.  Followup spectroscopic observations led to the measurement
of approximately 70 galaxy redshifts with the distribution seen in
Fig.~1.  The largest spike in the redshift distribution is at $z
\sim 0.63$, which is the redshift of the lensing galaxy.  To test
whether this spike represented a good group candidate, we examined the
spatial distribution of the galaxies in the spike.  The spike members
were centrally concentrated around the location of the lens (Fig.~2).
Therefore, we consider this as a likely group associated with the
lens.

\begin{figure}
 \includegraphics[height=3in,width=3in]{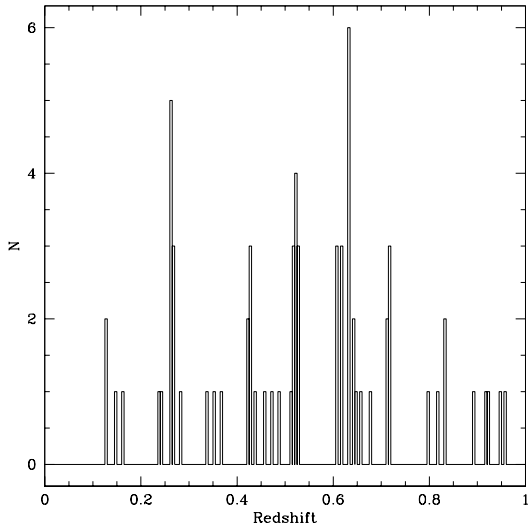}
  \caption{Redshift distribution in the field of B1608+656 (Fassnacht
   et al., in prep).}\label{fig:zhist}
\end{figure}

\begin{figure}
 \includegraphics[height=3in,width=3in]{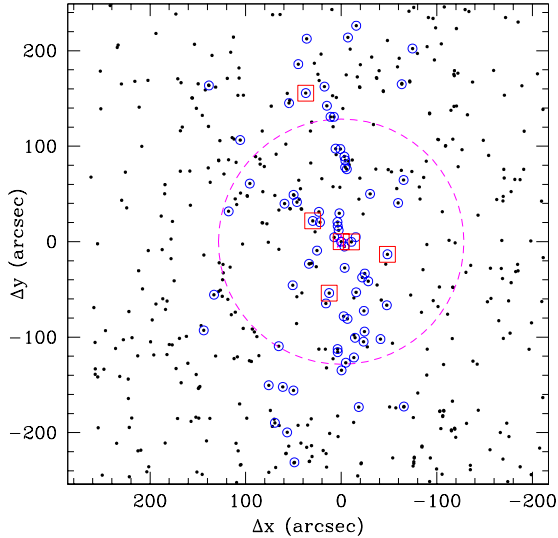}
  \caption{Spatial distribution of the galaxies in the B1608+656 field.
  Black dots represent galaxies with $r<23$, open circles represent
  galaxies with measured redshifts, and open boxes represent galaxies
  in the redshift spike.  The lens system is at the center of the
  plot at position (0,0).  The dashed circle has a radius of 1 $h^{-1}$
  comoving Mpc at the redshift of the lensing galaxy.  
  From Fassnacht et al., in prep.}\label{fig:xydist}
\end{figure}

To examine the effect of the group on the value of $H_0$ derived from
B1608+656 we must measure the group properties.  As a simple first
approximation, we describe the group as a singular isothermal sphere.
The velocity dispersion measured from the confirmed group members is
185$\pm$63~km/s, which corresponds to a convergence of $\kappa =
0.22/\theta$.  Here $\theta$ is the distance in arcseconds of the lens
system from the group centroid.  However, the value of $\theta$ is not
well determined because different measurements of the location of the
group centroid differ significantly.  The mean position of the group
galaxies and the luminosity-weighted position differ by more than 10
arcseconds, leading to a factor of $\sim$3 uncertainty in
$\kappa_{group}$ from the centroid position alone.  Finding a more
accurate group centroid will require significantly more spectroscopy
of the field, or a deep X-ray observation to detect the hot intragroup
gas.  However, the best approach may be to treat the group as a
collection of individual halos rather than one smooth overall
distribution, as suggested by the simulations of Keeton \& Zabludoff
(2004).  Once again, this will require more spectroscopic followup,
but a start can be made with the current data set (Fassnacht et al.,
in prep.).

\section{Other lens-group associations}

Through our dedicated survey we have discovered three additional galaxy
groups associated with gravitational lenses: a group in the foreground
of the CLASS B0712+472 system (Fassnacht \& Lubin 2002) and groups
associated with CLASS B2108+213 (McKean et al., in prep -- see
McKean poster at this meeting) and CLASS B1600+434 (Fassnacht et al.,
in prep).  Other spectroscopically-confirmed groups have been detected
in association with the lenses PG1115+080 (Kundi\'c et al.\ 1997a),
B1422+231 (Kundi\'c et al.\ 1997b; Tonry 1998), MG0751+2716 (Tonry
\& Kochanek 1999), and MG1113+0456 (Tonry \& Kochanek 2000).  This list
does not include groups discovered by the Zabludoff/Keeton
collaboration but not yet reported in the literature, nor the lens
systems associated with more massive galaxy clusters (e.g., Q0957+561,
RX J0911+0551, RX J0921+4529, SDSS J1004+4112; Young et al.\ 1980;
Kneib et al.\ 2000; Mu\~noz et al.\ 2001; Oguri et al.\ 2004), nor the
probable groups for which photometric redshifts exist but
spectroscopic confirmation has not yet been acquired (e.g., Rusin et
al. 2001; Faure et al.\ 2002).  Thus, the association of galaxy groups
and gravitational lenses provides a rich data set for not only
improving the determination of $H_0$, but also for studying group
properties over an interesting redshift range.


\newpage

\begin{discussion}

\discuss{Alloin}{If you had included groups in the southern sky your list of
lens-associated groups would have been longer.}

\discuss{Fassnacht}{I was including only groups which have
been spectroscopically confirmed.  However, I would not be surprised if I missed
some groups.  It would be unfortunate if they were all in the south.}

\discuss{Koopmans}{I notice that there are some gaps in the redshift
distribution for the B1608+656 field.  Do you want to say something about
those?}

\discuss{Fassnacht}{I know what you're getting at.  However, I don't think
that our spectroscopic sampling is dense enough to say anything meaningful
yet about the gaps.} 

\discuss{Kochanek}{You can't present the Keeton and Zabludoff simulation
results without mentioning the caveats.  After many emails with Chuck, I
have concluded that the width of the core in their distribution is just
due to the uncertainties in the time delays while the broad surrounding
distribution is due to the way that they did their simulations.}

\discuss{Fassnacht}{I didn't realize that.  However, I think that their main
point is still valid -- that if you don't include the groups your value of
$H_0$ will be systematically biased high.} 

\discuss{White}{Our simulations show that you can't approximate most groups
as a simple smooth distribution.  Instead they show multiple mass peaks.
Therefore, it would be better to model the group contribution as coming
from the individual halos.}

\discuss{Fassnacht}{I agree.  This is something that the Keeton and Zabludoff
simulations showed as well.  It is definitely something that needs to be
done for the B1608+656 group.} 

\end{discussion}

\end{document}